\documentclass{elsart}

\usepackage{epsfig}

\usepackage{amssymb}

\begin{document}

\begin{frontmatter}



\title{Superradiance in volume diffraction grating \thanksref{intas}}
\thanks[intas]{Authors thank INTAS for support (Grant ...)}


\author{V. Baryshevsky}, 
\author{K. Batrakov} 

\address{Research Institute of Nuclear Problem, 
Belarussian State University,   \\   11 Bobruyskaya Str. , Minsk 220050, Belarus}

\begin{abstract}
\qquad  To simulate VFEL operation  
the superradiance  from a short electron pulse moving in a volume diffraction 
grating is studied. It is supposed that Bragg condition for emitted photons is 
fulfilled and dynamical diffraction takes place. 
Spectral-angular distributions for transmitted and diffracted waves are derived. 
It is shown that the optimal geometry for superradiance exists and it is determined
by the energy and transverse size of electron beam.

\end{abstract}


\end{frontmatter}

\qquad 
\section{Introduction}
There are two ways to obtain coherent radiation: for unperturbed electron beam 
the coherent radiation appears as a result of bunching of electrons by ponderromotive wave. 
The second possibility exists at electron beam modulation or for short electron beam. 
If the size of electron beam is compared or less 
then  wavelength $l<\lambda$, the intensity of radiation is proportional to $N^2$, 
where $N$ is the number of electrons. As a result intensity of coherent radiation essentially exceeds the 
incoherent part (shot noise) in this case. Superradiance of electron beams was studied 
in a great number of works (see for example \cite{boni,jaro,ginz}). In present work 
the spectral-angular distributions of quasi-Cherenkov radiation in condition of volume distributed feedback for short electron beam is derived. 
Volume free electron lasers (VFEL) (\cite{volume}) use non-one-dimensional 
volume distributed feedback (VDFB), which essentually changes the dispersion characteristic 
of electromagnetic wave and retains radiation in interaction region. The sharp
increase of amplification process take place in the points of  degeneration 
(\cite{PhysRev,bar1}). The spectral-angular behavour of superradiance  in these point are modified also. 
      
\section{The spectral-angular distribution of quasi-Cherenkov superradiance}
The spectral-angular distribution of radiation is as follows \cite{bar}:
\begin{eqnarray}
W_{\mathbf{n}\omega }=\frac{e^{2}\omega ^{2}}{4\pi ^{2}c^{3}}\left|
\sum\limits_{i}\int dte^{i\omega t}\mathbf{v}_{i}(t)\mathbf{E}_{\mathbf{k}
}^{(-)\ast }(\mathbf{r}_{i}(t);\omega )\right| ^{2}. \label{distrib}
\end{eqnarray}
Here $\mathbf{E}_{\mathbf{k}}^{(-)}(\mathbf{r}_{i}(t);\omega )$ is the wave 
function of a photon, which asymptotically behaves as plane wave $\exp \{i\mathbf{kr}\}$ 
plus incoming spherical wave \cite{bar}, 
$\mathbf{v}_{i}(t)$ is the velocity of $i^{th}$ electron, $\omega$ and $\mathbf{k}$ are 
the frequency and wave vector of emitted photons.
The integration over whole electron pass
is performed in (\ref{distrib}).
Summing over all electrons in (\ref{distrib}) one can obtain the following form-factor 
of electron beam:
\begin{eqnarray}
\sum\limits_{ij}\exp
\{ i\mathbf{k}_{\perp }(\mathbf{r}_{j0\perp }-\mathbf{r}_{i0\perp })
+ik_{z}^{(ch)}(z_{i0}-z_{j0})\} =\label{form} \\
\int \int d\mathbf{r}_1d\mathbf{r}_2 n(\mathbf{r}_1)n(\mathbf{r}_2)
\exp\{ i\mathbf{k}_{\perp }(\mathbf{r}_{1\perp }-\mathbf{r}_{2\perp })
+ik_{z}^{(ch)}(z_{2}-z_{1})\}, \nonumber
\end{eqnarray}
where 
$\mathbf{r}_{i0}=(\mathbf{r}_{i0\perp },z_{i0})$ , 
$\mathbf{r}_{j0}=(\mathbf{r}_{j0\perp },z_{j0})$ are the coordinates of electrons at $t=0$, 
$n(\mathbf{r}_1)$ , $n(\mathbf{r}_2)$ are the microscopic density function 
$n(\mathbf{r})=\sum\limits_{i} \delta (\mathbf{r}-\mathbf{r}_{i0})$. 
Averaging (\ref{form}) over coordinates 
$\mathbf{r}_{i0}$ gives
\begin{eqnarray}
\overline{n(\mathbf{r}_{1})n(\mathbf{r}_{2})}=N\delta (\mathbf{r}_{1}-
\mathbf{r}_{2})+N(N-1)f^{(2)}(\mathbf{r}_{1},\mathbf{r}_{2}) \label{micro}
\end{eqnarray}
Here $f^{(2)}(\mathbf{r}_{1},\mathbf{r}_{2})$ is the two-particle distribution function. 
Neglecting corellation function one can write two-particle distribution function as
$f^{(2)}(\mathbf{r}_{1},\mathbf{r}_{2})=
f^{(1)}(\mathbf{r}_{1})f^{(1)}(\mathbf{r}_{2})$, 
where $f^{(1)}(\mathbf{r}_{1})$ is the one-particle distribution function. 
Substitution of (\ref{micro}) in (\ref{form}) gives:
\begin{eqnarray}
\overline{\int \int d\mathbf{r}_1d\mathbf{r}_2 n(\mathbf{r}_1)n(\mathbf{r}_2)
\exp\{ i\mathbf{k}_{\perp }(\mathbf{r}_{1\perp }-\mathbf{r}_{2\perp })+ik_{z}^{(ch)}(z_{2}-z_{1})\}}= \label{averag} \\
\left(  N+N(N-1)\left| \int d\mathbf{r} f(\mathbf{r})\exp\{ i\mathbf{k}^{(ch)}\mathbf{r}\}\right|^2\right) \nonumber 
\end{eqnarray}
Let us consider electron beam form-factors in some particular cases:
\begin{eqnarray}
for\  electron\  beam\  with\  rectangular\  cross\  section \nonumber \\ 
N+N(N-1)\left( \frac{\sin \{k_{x}\Delta x/2\}}{k_{x}\Delta x/2}\frac{\sin
\{k_{y}\Delta y/2\}}{k_{y}\Delta y/2}\frac{\sin \{k_{z}\Delta z/2\}}{%
k_{z}\Delta z/2}\right) ^{2}\nonumber \\ 
for\  electron\  beam\  with\  circular\  cross\  section \label{diff} \\ 
N+N(N-1)\left( \frac{\sin \{\omega l/2u\}}{\omega l/2u}\frac{%
J_{1}(k_{\perp }R)}{k_{\perp }R}\right) ^{2}\nonumber \\ 
for\  electron\  beam\  with\  Gauss\  profile\nonumber \\ 
N+N(N-1)\exp \{-k_{x}^{2}\frac{\sigma _{x}^{2}}{2}\}\exp \{-k_{y}^{2}\frac{%
\sigma _{y}^{2}}{2}\}\exp \{-k_{z}^{2}\frac{\sigma _{z}^{2}}{2}\} \nonumber
\end{eqnarray}
One can see from (\ref{diff}) that form-factor strongly depends on electron beam profile. 
For beam with Gaussian profile the  form-factor depends on frequency 
steadily. In case of sharp boundary of electron beam form-factor is the oscillating function
of frequency. 
Substituting (\ref{averag}) and the expression for $\mathbf{E}_{\mathbf{k}}^{(-)}(\mathbf{r}_{i}(t);\omega )$ in
(\ref{distrib}) one can obtain the following expression for
spectral-angular distribution of photons in Bragg diffraction geometry  
\cite{bar}:
\begin{eqnarray}
\frac{d^2 N_s}{d \omega d\Omega }=Q\frac {e^2  \omega }{4\pi^2
\hbar c^3}(\vec{e}_s \vec{v})^2 \times \label{direct} \\
 \left| \sum_{\mu =1,2}\gamma _{\mu s}^{0} e^{ i\frac{\omega }{c\gamma _{0}}
\varepsilon _{\mu s}L}\left[ \frac{1}{\omega -\vec{k}\vec{v}}-\frac{1}
{\omega -\vec{k}_{\mu s}\vec{v}}\right] 
\left[ e^{\frac{i(\omega -\vec{k}_{\mu s}\vec{v})L}
{c\gamma _{0}}}-1\right] \right| ^{2}, \nonumber
\end{eqnarray}
The denotions introducine in \cite{bar} is used in (\ref{direct}), $Q$ is the form-factor.
For photons with the polarization vector $\vec{e}_{\tau s}$ emitted at
diffraction direction  can be rewritten as follows \cite{bar}:
\begin{eqnarray}
\frac{d^{2}N_{s}}{d\omega d\Omega }=Q_\tau \frac{e^{2}\omega }{4\pi
^{2}\hbar c^{3}}(\vec{e}_{\tau s}\vec{v})^{2} \label{diffracted}\\
\left| \sum_{\mu =1,2}\gamma _{\mu s}^{\tau }\left[ \frac{1}{\omega -\vec{k}_{\tau }
\vec{v}}-\frac{1}{\omega -\vec{k}_{\mu \tau s}\vec{v}}\right] \left[
e^{\frac{i(\omega -\vec{k}_{\mu \tau s}\vec{v})L}{c\gamma _{0}}}-1\right]
\right| ^{2}.\nonumber
\end{eqnarray}
In (\ref{diffracted}) $Q_\tau$ is the form-factor for diffracted photons:
\begin{eqnarray}
Q_\tau=  
 N+N(N-1)\left|\int d\mathbf{r} f(\mathbf{r})\exp\{ i\mathbf{k}^{(ch)}_\tau \mathbf{r}\}\right|^2 \label{formdif} 
\end{eqnarray}
The diffraction anomalies appear in 
spectral-angular distributions of spontaneous radiation near the point of root degeneration  
in conditions of dynamical diffraction \cite{batr1}. 
Two factors arouse these peaks. 
The first one is the decreasing of the wave group velocity:
\begin{eqnarray}
v^{(gr)}_z=\frac{2\gamma_0 c}{1+\beta \mp \frac{\Delta}{\varkappa}(1-\beta)},\label{groupvel}
\end{eqnarray}
where $\Delta=-\chi _0(1-\beta) -\beta \alpha$ , $\varkappa=\sqrt{\Delta^2+4\beta r}$.
The second condition for apearance of diffraction spectral-angular anomaly 
is the phase condition:
\begin{eqnarray}
(k_{1z}-k_{2z})L=2\pi n, \label{phase}
\end{eqnarray}
where $n$ is integer, $k_{1z}$ and $k_{2z}$ are the roots of dispersion equation 
for two wave dynamical diffraction. 
In these conditions the additional factor $(k\chi L/4\pi)^2 \gg 1$ 
appears in expressions for spectral-angular density. 
The remarkable property of points of diffraction describing by (\ref{phase}) 
is the fact that (\ref{phase}) corresponds to maximal radiation intensity and 
maximal value of amplification due to stronger interaction of photons with 
spartially periodic medium. The same conditions (\ref{groupvel}), (\ref{phase}) correspond to more intensive interaction in two wave VFEL (\cite{bar1})

\begin{figure}[h]
\epsfxsize = 12 cm 
\centerline{\epsfbox{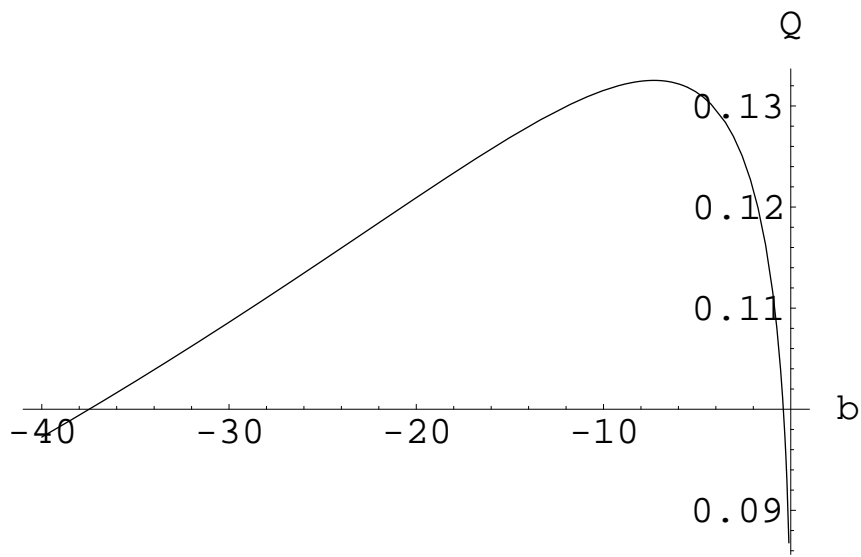}}
\caption{Dependence of form-factor of quasi-Cherenkov superradiation on 
asymmetry factor $\beta$.}
\label{asymm}
\end{figure}

There are optimal parameteres of diffraction geometry, for which the spectral-angular 
density of photon in the superradiance process increases. 
The dependence of form-factor on asymmetry factor 
$\beta=\gamma _0/\gamma_1$ is presented on fig.\ref{asymm}.

The optimal geometry is determined by many factors such as electron beam profile and energy, 
electrodynamical properties of grating, electron bunch length and so on. 

\section{Conclusion}
Quasi-Cherenkov superradiance in the
volume  diffraction grating can be obtained 
at the absence of Cherenkov superradiance, it is possible if:\\
1) Cherenkov condition for radiation is not fulfilled, but quasi-cherenkov condition 
is satisfied \cite{bar} or \\
2) cross section of electron beam meets the condition
$k\Theta_{ch} R \gg 1$, where $R$ is electron beam radius. Then there is possibility to 
observe superradiance at angles lower than $\Theta_{ch}$ for quasi-Cherenkov radiation 
mechanism due to equality  
$$
\Theta_{q-ch}^2=\Theta_{ch}^2+\frac{-\alpha \pm \sqrt{\alpha^2+4 r}}{2}
$$
one can obtain superradiance at range of Bragg parameteres corresponding to angles $k\Theta_{q_ch} R \sim 1$
The similarity between quasi-Cherenkov lasing process and quasi-Cherenkov radiation from  short electron beam
allows the experimental simulation  of induced radiation in condition of VDFB. For example by using the volume diffraction grating \cite{bar4} with spatial period of about few milllimeters we can study the quasi Cherenkov superradiance, to check the dependence of quasi Cherenkov radiation on parameter $\beta$ and to observe  the spectral angular diffraction anomalies corresponding to (\ref{groupvel}) and (\ref{phase}) in millimeter wavelength range. For volume grating with length $L=10$ cm the angular and spectral resolution 
$\Delta \Theta \sim \Delta \omega /\omega \sim 10^{-3}$ for observing of diffraction anomalies is needed.


\end{document}